\documentclass[final,5p,times,twocolumn]{elsarticle}
\biboptions{square,sort&compress,comma,numbers}
\pdfoutput=1

\usepackage[latin1]{inputenc}
\usepackage[english]{babel}
\usepackage{amssymb,amsmath,amsthm,cancel,graphicx,xcolor}
\usepackage{booktabs}
\usepackage{mathrsfs}

\newcommand{\abs}[1]{\left| #1 \right|}

\let\originalleft\left
\let\originalright\right
\renewcommand{\left}{\mathopen{}\mathclose\bgroup\originalleft}
\renewcommand{\right}{\aftergroup\egroup\originalright}

\def\beq{\begin{equation}}  
\def\eeq{\end{equation}}
\def\({\left(}
\def\){\right)}
\def\[{\left[}
\def\]{\right]}

\def\eq#1{{Eq.~(\ref{#1})}}
\def\eqs#1#2{{Eqs.~(\ref{#1})--(\ref{#2})}}
\def\fig#1{{Fig.~\ref{#1}}}

\def\Table#1{{Table~\ref{#1}}}

\def\sect#1{{Sect.~\ref{#1}}}

\def\abs#1{\left| #1\right|}

\def\det{\mbox{det}\,}

\journal{arXiv}

\begin{document}

\begin{frontmatter}



\title{Stability of the electroweak ground state in the Standard Model
and its extensions}


\author[label1]{Luca Di Luzio} 
\author[label2]{Gino Isidori} 
\author[label1]{Giovanni Ridolfi} 
\address[label1]{Dipartimento di Fisica, Universit\`a di  Genova and INFN,
Sezione di Genova, Via Dodecaneso 33, I-16146 Genova, Italy}
\address[label2]{Department of Physics,
University of Z\"urich, Winterthurerstrasse 190,
CH-8057 Z\"urich, Switzerland}

\begin{abstract}
We review the formalism by which the tunnelling probability of an
unstable ground state can be computed in quantum field theory, with
special reference to the Standard Model of electroweak
interactions. We describe in some detail the approximations implicitly
adopted in such calculation. Particular attention is devoted to the
role of scale invariance, and to the different implications of
scale-invariance violations due to quantum effects and possible new
degrees of freedom.  We show that new interactions characterized by a
new energy scale, close to the Planck mass, do not invalidate the main
conclusions about the stability of the Standard Model ground state
derived in absence of such terms.

\end{abstract}


\end{frontmatter}


\section{Introduction}
\label{intro}

In recent years there has been considerable interest in the problem of
the stability of the Standard Model (SM) ground state.  Due to the
sizable negative contribution to the $\beta$ function of the Higgs
self-coupling induced by top-quark loops, the usual
electroweak vacuum $|0\rangle$, characterized by
$\langle 0 | h | 0 \rangle = v \approx 246$~GeV, may not be the
absolute minimum of the scalar potential. 
In this case the true
minimum of the theory is located at much larger energy scales and, in
the absence of New Physics (NP) modifying the effective Higgs potential,
the electroweak vacuum is unstable. The parameter space of the SM
(with particular reference to the top-quark mass, $m_t$, and the
Higgs-boson mass, $m_h$, which are the most relevant parameters) is
thus naturally divided into three regions: stability, instability, and
metastability.  The stability region is the one where the electroweak
vacuum is the absolute minimum of the potential. The instability and
metastability regions are those where a new deeper minimum exists,
with the metastability region being characterized by a lifetime of the
unstable electroweak vacuum larger than the age of the Universe.  More
precisely, the instability/mestastability boundary is determined by
the decay probability of the electroweak vacuum under quantum
tunnelling, that sets a model-independent upper bound on the lifetime
of the unstable vacuum irrespective of the thermal history of the
Universe.

A precise determination of the boundaries of these three regions has
recently been presented in
Refs.~\cite{Degrassi:2012ry,Buttazzo:2013uya} 
(see also \cite{Bezrukov:2012sa,Alekhin:2012py,Bednyakov:2015sca}).  As a result of these
recent analyses, the present experimental values of $m_h$ and $m_t$ lie in the
metastability region of the SM parameter space. This finding holds
only within the SM; however, it has an important consequence for
beyond-the-SM searches: it implies that there is no need to invoke the
presence of NP in order to stabilize the SM electroweak
vacuum.

The validity of the analysis in
Refs.~\cite{Degrassi:2012ry,Buttazzo:2013uya} and the corresponding
conclusions has been questioned in a series of recent
papers~\cite{Branchina:2013jra,Branchina:2014usa,Branchina:2014rva,Branchina:2015nda}.
There, it has been shown that non-standard physics modifying the
shape of the Higgs potential at energy scales of the order of the
Planck mass can sizeably affect the tunnelling rate of the electroweak
vacuum.  This observation is correct. However, as we discuss in the
following, it does not invalidate the interest and the main
conclusions of the analyses based on the SM potential.

In this paper we present a critical re-analysis of the problem of the
SM vacuum stability.  Our purpose is to clarify the assumptions and
approximations employed in the evaluation of the SM tunnelling rate,
with particular attention to those that have been often overlooked or
implicitly adopted in the existing literature.

\section{The tunnelling rate within the SM}

We begin by reviewing the standard formalism, originally worked out by
Coleman and Callan~\cite{Coleman:1977py,Callan:1977pt}, which allows
one to compute the probability per unit time (or, equivalently, the
lifetime) of a false ground state to a true ground state in quantum
field theory. In the semiclassical approximation, the decay
probability per unit time of the electroweak ground state is given
by~\cite{Isidori:2001bm}
\beq \Gamma\approx \frac{\tau_U^3}{R^4}e^{-S[h]},
\label{rate}
\eeq
where $\tau_U$ is the age of the Universe,
\beq
\label{defaction}
S[h]=\int d^4x\,\[\frac{1}{2}\partial_\mu h\partial_\mu h
+V(h)\]
\eeq
is the euclidean action of the theory, computed for a specific
solution $h$ of the euclidean field equation for the scalar field
which is usually called the {\it bounce}, and $R$ is a dimensional
factor associated with the size of the bounce.  The bounce field
configuration is such that it is equal to the false vacuum
configuration $h=v$ at infinite euclidean time $\tau$, and completes
barrier penetration at $\tau=0$.

A known result, conjectured by Coleman and subsequently proved by
Coleman himself, Glaser and Martin~\cite{Coleman:1977th}, guarantees
that the bounce solution of minimum action is invariant under
four-dimensional rotations in euclidean spacetime, that is
\beq
h=h(r);\qquad r^2=|\vec x|^2+\tau^2.
\eeq
Hence,
\beq
\lim_{r\to\infty}h(r)=v.
\label{boundary1}
\eeq
By requiring that the solution is non-singular at the origin, we also have
\beq
\left.\frac{dh(r)}{dr}\right|_{r=0}=0.
\label{boundary2}
\eeq
Eq.~(\ref{rate}) gives the leading contribution to the tunnelling rate
in the semiclassical limit, that is, it only includes exponentially
enhanced terms in the limit $\hbar\to 0$. In particular, the overall
normalization can only be determined by including the first quantum
corrections~\cite{Isidori:2001bm}.

We observe that the bounce is the unique solution of a suitably
defined Cauchy problem. Indeed, under the assumption that
the bounce is $O(4)$-invariant, the field equation
for the bounce is
\beq
h''(r)=-\frac{3}{r}h'(r)+V'(h),
\label{bounce}
\eeq
where $V(h)$ is the scalar potential, and primes denote differentiation
with respect to the functional argument. A set of initial conditions
\beq
h(r_0)=h_0;\qquad h'(r_0)=h_1,
\eeq
given at any finite value $r_0\ne 0$, defines a Cauchy problem, which
has a unique solution in a neighbourhood of $r_0$, as a consequence of
known results in real analysis. Since $r=0$ is the only singular point
in the r.h.s.~of Eq.~(\ref{bounce}), $r_0$ can be taken to be
arbitrarily large. However, the unique solution is not necessarily
well defined at $r=0$, because of the singularity in the r.h.s.~of
Eq.~(\ref{bounce}), nor at $r\to\infty$, because the existence and
unicity theorem has a local meaning. Thus, for a generic choice of the
initial conditions at $r=r_0$, the boundary conditions
Eqs.~(\ref{boundary1},\ref{boundary2}), are not necessarily
fulfilled. Conversely, the requirement that
Eqs.~(\ref{boundary1},\ref{boundary2}) are fulfilled by the bounce
selects a set of allowed initial conditions at an intermediate point
$r_0$.  Depending on the shape of the potential, it is possible to
have multiple bounce solutions (with different initial values $h_0$
and $h_1$) which satisfy the boundary conditions in
Eqs.~(\ref{boundary1},\ref{boundary2}).

\subsection{Decay of the SM vacuum in the semiclassical
approximation}

The case of the pure SM is especially interesting. In this case, the
value of the true vacuum is typically very large with respect to the
electroweak scale $v\approx 246$~GeV, so one usually takes $v=0$. Thus
\beq
\lim_{r\to\infty}h(r)=0.
\label{boundary1sm}
\eeq
The validity of this approximation is discussed below in Sect.~\ref{mneq0}.
The scalar potential in the unstable region is therefore 
\beq
V(h)=\frac{1}{4}\lambda h^4,
\label{SMpot}
\eeq
where we neglect the logarithmic running of $\lambda$ 
and take it as a negative constant. 
This approximation is reviewed in \sect{violscalradiative}. 
Then \eq{bounce} takes the form
\beq
h''(r)+\frac{3}{r}h'(r)=\lambda h^3(r).
\label{bouncexSM}
\eeq
Eq.~(\ref{bouncexSM}) is invariant under scale transformations:
if $h(r)$ is a solution,
then
\beq
h_a(r)=ah(ar)
\label{scale}
\eeq
is also a solution, for any choice of the scale factor $a$. Indeed
\begin{multline}
h_a''(r)+\frac{3}{r}h_a'(r)=a^3\left[h''(ar)+\frac{3}{ar}h'(ar)\right] \\
=\lambda a^3h^3(ar)=\lambda h_a^3(r).
\end{multline}
Obviously, the scaled solution
$h_a(r)$ has the same limiting behaviors Eqs.~(\ref{boundary1},
\ref{boundary2}) as the original one, 
but different initial conditions at $r=r_0$. Otherwise stated, the
boundary conditions do not fix the overall normalization of the bounce
solution.

As is well known, a solution of Eq.~(\ref{bouncexSM}) with the
boundary conditions (\ref{boundary1},\ref{boundary2}) is given by
the Fubini-Lipatov instanton~\cite{Fubini:1976jm,Lipatov:1976ny}
\beq
h(r)=\sqrt{\frac{8}{|\lambda|}}\frac{R}{R^2+r^2},
\label{SMh}
\eeq
for any value of $R$ and $\lambda<0$ (note that $h(R)=h(0)/2$: this
will be our definition of the size of the bounce throughout the paper).
It should be clear from the
above discussion that the presence of the arbitrary parameter $R$ is
just a reflection of the scale invariance of the equation, and not a
signal of non-unicity of the solution; indeed, the scaling defined in
Eq.~(\ref{scale}) amounts to replacing $R$ with $R/a$ in
Eq.~(\ref{SMh}).  $h(r)$ is the unique solution of Eq.~(\ref{bounce})
with initial conditions in $r_0$
\beq
h_0=\sqrt{\frac{8}{|\lambda|}}\frac{R}{R^2+r_0^2};\qquad
h_1=-\sqrt{\frac{8}{|\lambda|}}\frac{2Rr_0}{(R^2+r_0^2)^2},
\eeq
which obey the constraints (\ref{boundary1},\ref{boundary2}).

The SM bounce Eq.~(\ref{SMh}) can be found by the following procedure
(see e.g.~Ref.~\cite{Lee:1985uv} for an alternative derivation).
Let us assume that a solution of Eq.~(\ref{bouncexSM}) exists,
with a Taylor expansion around $r=0$:
\beq
h(r)=\sum_{k=0}^\infty A_k r^k
\label{bounceTaylor}
\eeq
with $A_0>0$. Eq.~(\ref{bouncexSM}) takes the form
\beq
\frac{3A_1}{r}+
\sum_{k=0}^\infty (k+2)(k+4)A_{k+2}r^k
=\lambda\sum_{k=0}^\infty r^k\sum_{i=0}^k\sum_{j=0}^{k-i}A_i A_j A_{k-i-j}.
\eeq
It follows that
\beq
A_1=0.
\eeq
Thus, the condition that the first derivative of the bounce at $r=0$ vanishes
is a consequence of the assumption Eq.~(\ref{bounceTaylor}).
The remaining coefficients are given by the recurrence relation
\beq
A_{k+2}=\frac{\lambda}{(k+2)(k+4)}\sum_{i=0}^k\sum_{j=0}^{k-i}A_i A_j A_{k-i-j}.
\eeq
The coefficients $A_k$ with $k$ odd are zero: indeed, when $k$ is odd,
one out of the three summation indices $i,j,k-i-j$ is also odd 
(possibly all of them).
Hence $A_1=0$ implies $A_3=0$, and so on.
Thus, we may rewrite Eq.~(\ref{bounceTaylor}) as
\beq
h(r)=\sum_{k=0}^\infty a_k r^{2k};\qquad a_k=A_{2k},
\label{bounceTayloreven}
\eeq
and the recurrence relation for the coefficients becomes
\beq
a_{k+1}=\frac{\lambda}{8}\frac{2}{(k+1)(k+2)}
\sum_{i=0}^k\sum_{j=0}^{k-i}a_i a_j a_{k-i-j}.
\label{recurrence}
\eeq
The coefficients $a_k$ are determined by the single number $a_0$, the
value of the bounce at the origin. We now show that
\beq
a_j=\left(\frac{\lambda}{8}\right)^j a_0^{2j+1}.
\label{aj}
\eeq
The proof is by induction. For $k=0$ Eq.~(\ref{recurrence})
gives
\beq
a_1=\frac{\lambda}{8}a_0^3.
\eeq
We now assume that Eq.~(\ref{aj}) holds for $0\leq j\leq k$. Then
\begin{align}
a_{k+1}=&\frac{\lambda}{8}\frac{2}{(k+1)(k+2)}
\sum_{i=0}^k\sum_{j=0}^{k-i}a_i a_j a_{k-i-j}
\nonumber\\
=&\frac{\lambda}{8}\frac{2}{(k+1)(k+2)}
\sum_{i=0}^k\sum_{j=0}^{k-i}\left(\frac{\lambda}{8}\right)^k a_0^{2k+3}
\nonumber\\
=&\left(\frac{\lambda}{8}\right)^{k+1} a_0^{2k+3}
\end{align}
which is what we set out to prove. The Taylor expansion in 
Eq.~(\ref{bounceTayloreven}) can now be summed. We find
\beq
h(r)=a_0\sum_{k=0}^\infty \left(\frac{\lambda}{8}\right)^k a_0^{2k}r^{2k}
=\frac{a_0}{1-\frac{\lambda}{8}a_0^2r^2}.
\label{bounceSM2}
\eeq
The series has convergence radius
\beq
R=\sqrt{\frac{8}{|\lambda|}}\frac{1}{a_0},
\label{Rdef}
\eeq
but if $\lambda$ is negative the sum can be analytically continued to
the whole positive real axis, and vanishes as $r\to\infty$. Finally, we note
that for $\lambda<0$ the bounce in Eq.~(\ref{bounceSM2}) coincides
with the solution given in Eq.~(\ref{SMh}), with $R$ as in
Eq.~(\ref{Rdef}). The above construction shows that, given the value
of the bounce at the origin and the requirement of regularity on the
range $0<r<\infty$, the solution is unique.

The value of the euclidean  action $S[h]$ of the bounce
solutions Eq.~(\ref{SMh}) is
\beq
\label{BactionSM}
S[h]=\frac{8\pi^2}{3|\lambda|},
\eeq
independently of the value of $R$. This is not surprising,
because the action is 
dimensionless in natural units, and no dimensionfull scale parameter is available
other than $R$. Hence, there is no way to single out one preferred
value of $R$ at the semiclassical level.
However, $R$ is related to the value of the bounce 
at $r=0$:
\beq
h(0)=\sqrt{\frac{8}{|\lambda|}}\frac{1}{R},
\eeq
and since the bounce solution only exists for $\lambda<0$, we expect that
\beq
\frac{1}{R} > \Lambda_I,
\eeq
where $\Lambda_I\approx 10^{10}$~GeV is the energy scale at which the running
coupling $\lambda(\mu)$ becomes negative.
This is an {\it a posteriori} confirmation that neglecting
the electroweak scale, of order $10^2$~GeV, with
respect to the size of the bounce is indeed a reliable approximation.

\subsection{Violation of scale invariance through radiative corrections}
\label{violscalradiative}

The first quantum corrections, computed in Ref.~\cite{Isidori:2001bm},
affect the semiclassical result in two respects: they fix the
normalization in Eq.~(\ref{rate}), and they take into account the
running of the Higgs coupling $\lambda$.\footnote{For some
    earlier works discussing the breaking of scale invariance due to
    the running of $\lambda$ in the context of the SM-vacuum
    tunnelling calculation see also
    Refs.~\cite{Arnold:1989cb,Arnold:1991cv}.} As a consequence, the
tunnelling decay rate is dominated by the bounce with the maximum
value of $|\lambda(\mu)|\sim|\lambda(1/R_{\rm SM})|$. For the central
values of the SM parameters the scale $1/R_{\rm SM}$ turns out to be a
couple of order of magnitudes below the Planck mass $M_P= 1.22 \times
10^{19}$~GeV.

Schematically, one expands the euclidean action around the tree-level
bounce solution
\beq
S[h+\tilde{h}] \approx S[h] + \frac{1}{2} 
\int d^4 x \, S^{''} [h] \, \tilde h^2,
\eeq
and integrates over the fluctuations $\tilde h$:
\beq
\int \mathcal{D} [\tilde h] e^{-S[h+\tilde h]} 
\approx  e^{-S[h]} \( \det S^{''} [h] \)^{-\frac{1}{2}}.
\eeq
In the treatment of the functional determinant there are two main
aspects which eventually lead to the appearance of the scale $\mu \sim
1/R_{\rm SM}$ in the calculation: $i)$ ultraviolet (UV) divergences of the non-zero
modes of $S^{''} [h]$, responsible for the introduction of a
renormalization scale $\mu$ and $ii)$ the treatment of the zero modes
of $S^{''} [h]$, which have to be singled out and treated separately
in order to avoid unphysical divergences.  The existence of zero modes
is simply a reflection of the fact that the classical action is
invariant under a larger class of symmetries which are broken by the
explicit solution of the tree-level bounce (e.g.~in the case of
\eq{SMh} these are translations in $O(4)$ and scale transformations).
The symmetries of the action are hence restored only if one considers
the family of all bounce solutions as a whole.  This defines a measure
of integration (the instanton measure $d \mu_{\rm inst}$~\cite{'tHooft:1976fv})
over the
the so-called collective coordinates, parametrizing the families of
bounces with different sizes ($R$) and located anywhere in euclidean
spacetime ($x_0$)
\beq
d \mu_{\rm inst} \approx 
\frac{d R d^4 x_0}{R^5} \exp \[ -\frac{8 \pi^2}{3 \abs{\lambda (1/R)}} 
- \Delta S \], 
\eeq
where $\lambda (1/R)$ is the renormalized coupling at the scale $1/R$,
$\Delta S = \mathcal{O}\(g_i^2 / \lambda\)$ (with $i$ running over the
SM couplings) comes from the inclusion of the non-zero modes and
contains finite terms plus logs which are minimized by the choice $\mu
\sim 1/R$ \cite{Isidori:2001bm}.  The tunnelling probability is
therefore obtained by integrating over the collective coordinates
\beq
\label{pintmu}
p = \int d \mu_{\rm inst} \approx \tau^4_U \int \frac{d R}{R^5} 
\exp \[ -\frac{8 \pi^2}{3 \abs{\lambda (1/R)}} - \Delta S \], 
\eeq
where the
integral is extended to the range where $\lambda (1/R)<0$,
where a bounce exists.

By expanding $\lambda(1/R) < 0$ around its minimal value located at $R = R_{\rm SM}$
\beq
\lambda (1/R) \approx \lambda (1/R_{\rm SM}) 
+ \frac{1}{2} \lambda'' (1/R_{\rm SM}) (1/R - 1/R_{\rm SM})^2, 
\label{1overRexp}
\eeq
we can approximate the integral by the method of the steepest descent
\begin{multline}
p \approx \frac{\tau^4_U}{R_{\rm SM}^3} 
\sqrt{\frac{3 \lambda^2(1/R_{\rm SM}) }{4 \pi \lambda'' (1/R_{\rm SM})}} 
\exp\[-\frac{8\pi^2}{3 | \lambda(1/R_{\rm SM}) | } - \Delta S \] \\
\approx \frac{\tau^4_U}{R_{\rm SM}^4} \exp\[-\frac{8\pi^2}{3 | \lambda(1/R_{\rm SM}) | }\], 
\label{papproxSM}
\end{multline}
where in the last step we have neglected the subleading pre-exponential
factors (including $\Delta S$), and we have set
$\lambda'' (1/R_{\rm SM}) \sim R_{\rm SM}^2$ on dimensional grounds.

\begin{figure}[ht]
  \begin{center}
    \includegraphics[width=.45\textwidth]{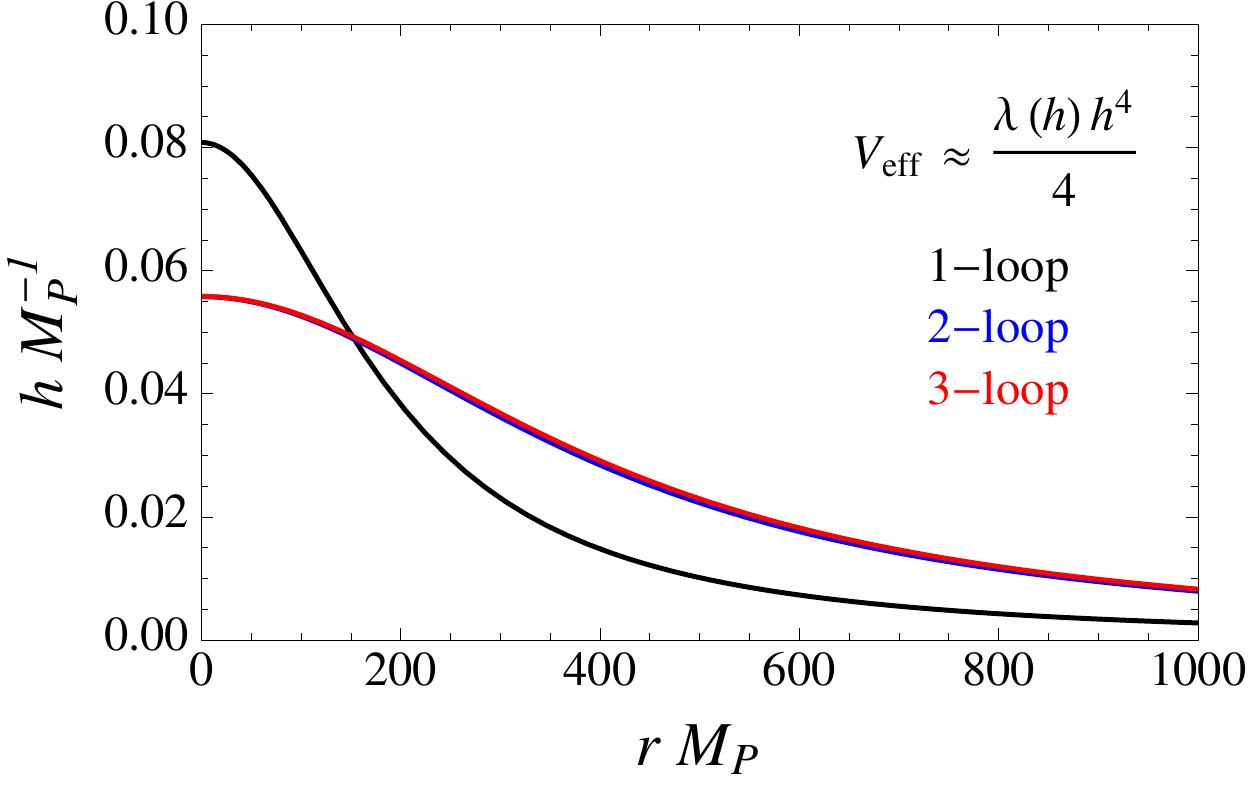}
  \end{center}
\caption{\label{Bounce_lamrun}
  Bounce profiles for the potential in \eq{SMtreepotimproved}, where
  the running of $\lambda$ is evaluated at one (black),
  two (blue) and three (red) loops, respectively. 
  The adimensional variables are normalized in terms of the Planck mass
  $M_P$.}
\end{figure}

\begin{figure}[ht]
  \begin{center}
    \includegraphics[width=.45\textwidth]{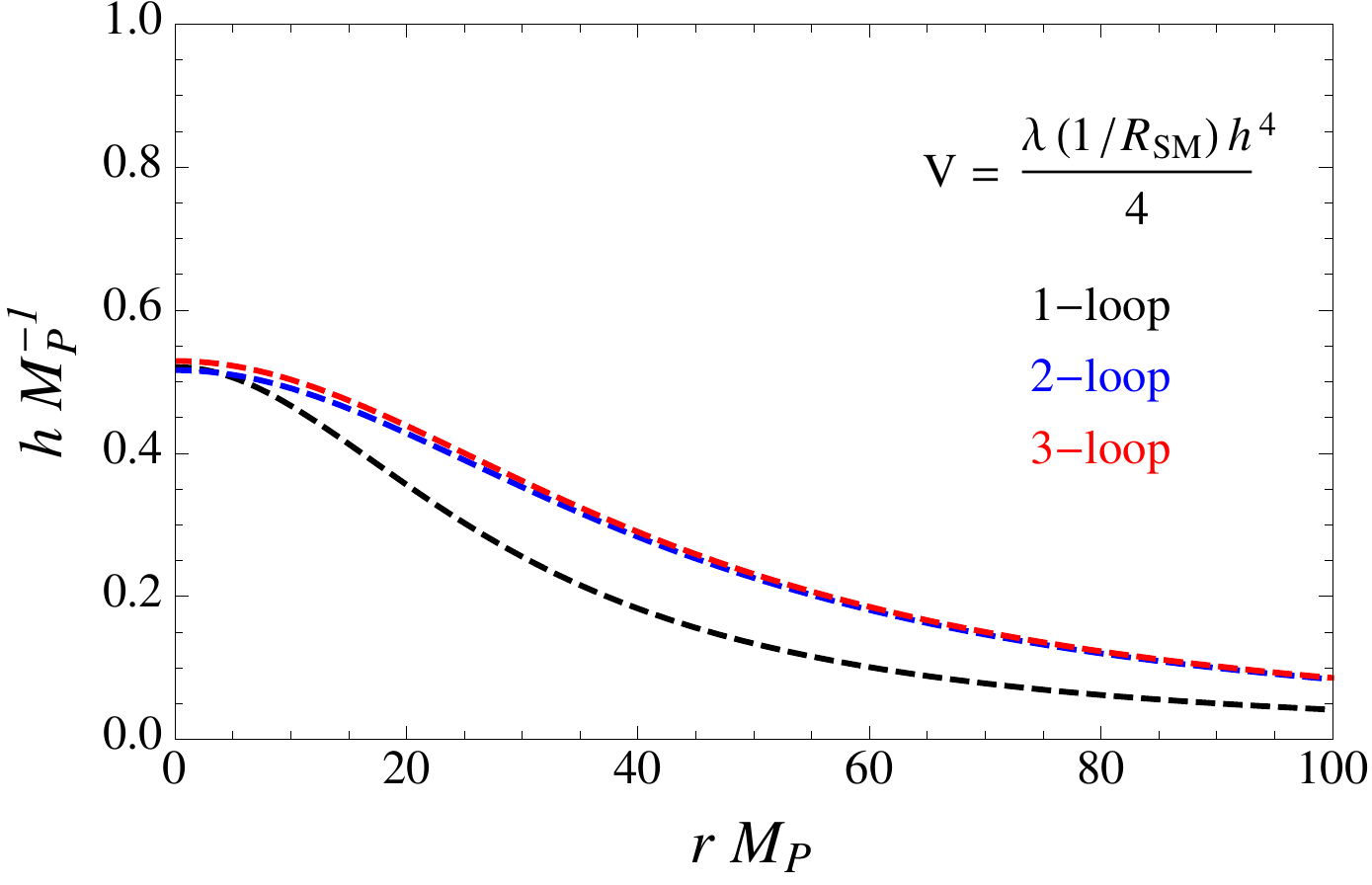}
  \end{center}
\caption{\label{Bounce_lammin} 
 Same as in \fig{Bounce_lamrun} for the SM tree-level potential with 
 $\lambda (1/R_{\rm SM})$ a negative constant.}
\end{figure}

The result \eq{papproxSM} provides the standard approximation which is usually
employed in order to estimate the lifetime of the electroweak
vacuum. Nonetheless, it is interesting to compare it with a more
direct calculation obtained via the renormalization group (RG) improved effective
potential
\beq
V_{\rm eff}=\frac{1}{4} \lambda_{\rm eff} (h) h^4 \approx \frac{1}{4} \lambda (h) h^4, 
\label{SMtreepotimproved}
\eeq
where, as in \eq{SMpot}, we neglected the mass term and in the last step  
we have approximated the effective 
quartic coupling $\lambda_{\rm eff}$ 
with the $\overline{\text{MS}}$ renormalized
coupling $\lambda$ evaluated at $\mu = h$. 
The scale ambiguity of the
tree-level bounce is now resolved by the leading order effective
potential which takes into account the dominant radiative corrections.
However, since it is not possible to find an analytical solution for
the bounce one has to resort to a numerical analysis (see
\ref{numerical} for details).  By applying our numerical
set-up\footnote{In this work we refer to arXiv v4 of
  \cite{Buttazzo:2013uya} for the central values of the SM couplings
  evaluated at the top-quark pole-mass scale.  In particular, we take
  $m_t = 173.34$ GeV and $m_h = 125.15$ GeV.}  we obtain the bounces
displayed in~\fig{Bounce_lamrun} where, for illustrative purposes, we
considered the case where the running of $\lambda$ is evaluated at
one (${\rm 1\ell}$), two (${\rm 2\ell}$) and three (${\rm 3\ell}$) loops, respectively.
The corresponding bounce actions are found to be 
\begin{align}
\label{Bnew1l}
S[h,\lambda(h)^{\rm 1\ell}] &= 772.3,  \\
\label{Bnew2l}
S[h,\lambda(h)^{\rm 2\ell}] &= 1703.9,  \\
\label{Bnew3l}
S[h,\lambda(h)^{\rm 3\ell}] &= 1788.8.
\end{align}
On the other hand, we want to compare the latter case with the 
standard method (cf.~\eq{papproxSM}) where $\lambda$ is evaluated 
at the scale $1/R_{\rm SM}$ where 
$\beta_\lambda \equiv \frac{d \lambda}{d \log \mu} = 0$. 
For our set of parameters, we get\footnote{By taking into account the 3$\sigma$ error bands of the most important 
SM parameters there is about one order of magnitude uncertainty on the scale where 
$\beta_\lambda=0$ \cite{Buttazzo:2013uya}.} 
\begin{eqnarray}
\!\!\!\!\!\! 1/R_{\rm SM}^{\rm 1\ell} = 4.145 \times 10^{17} \ \text{GeV}; &
\lambda (1/R_{\rm SM})^{\rm 1\ell} = - 0.03409,  \nonumber \\ 
\!\!\!\!\!\! 1/R_{\rm SM}^{\rm 2\ell} = 2.768 \times 10^{17} \ \text{GeV}; &
\lambda (1/R_{\rm SM})^{\rm 2\ell} = - 0.01546, \nonumber \\ 
\!\!\!\!\!\! 1/R_{\rm SM}^{\rm 3\ell} = 2.769 \times 10^{17} \ \text{GeV}; &
\lambda (1/R_{\rm SM})^{\rm 3\ell} = - 0.01473, \nonumber
\end{eqnarray}
which yields the bounces displayed in \fig{Bounce_lammin}. 
Correspondingly, the bounce actions defined by 
\begin{equation}
\label{Bstandard}
S[h,\lambda (1/R_{\rm SM})] = \frac{8 \pi^2}{3 |\lambda (1/R_{\rm SM})|} 
\end{equation}
have the following values
\begin{align}
\label{Bst1l}
S[h,\lambda (1/R_{\rm SM})^{\rm 1\ell}] &= 772.0, \\ 
\label{Bst2l}
S[h,\lambda (1/R_{\rm SM})^{\rm 2\ell}] &= 1702.6, \\ 
\label{Bst3l}
S[h,\lambda (1/R_{\rm SM})^{\rm 3\ell}] &= 1787.4.
\end{align}
Although the profiles of the bounces in \fig{Bounce_lamrun} and
\fig{Bounce_lammin} look different, the bounce actions in
\eqs{Bnew1l}{Bnew3l} and \eqs{Bst1l}{Bst3l} are very
similar.\footnote{This fact can be understood as follows: the bounces
  in \fig{Bounce_lammin} can be rescaled by $h(r) \rightarrow
  ah(ar)$ in such a way that they almost superimpose with those in
  \fig{Bounce_lamrun}.  The corresponding change in the bounce action,
  induced by the rescaling $R_{\rm SM}\to R_{\rm SM}/a$, is small, because
  $\lambda$ varies very slowly around its minimum.}  In particular,
\eq{Bstandard} provides a lower bound on the bounce action, since it
is obtained by expanding $\lambda(1/R)$ around its minimum.

To conclude, the good agreement between the two procedures 
for the determination of the decay probability 
justifies the approximation made in \cite{Isidori:2001bm}
of taking a constant $\lambda < 0$ for the leading order SM bounce.  
This was not completely obvious a priori, since computing the bounce 
``does not commute'' with the running of $\lambda$. 
Of course, the potential with running $\lambda$ in \eq{SMtreepotimproved} 
only captures log-enhanced corrections to the tunnelling rate, 
while a complete one-loop calculation requires 
the determination of the SM action functional around the 
leading order bounce configuration \cite{Isidori:2001bm}.

\subsection{Gauge independence of the tunnelling rate}

A question which is directly related to the calculation of quantum corrections 
to the SM vacuum decay rate is that of gauge invariance. 
Indeed, if one naively takes into account loop corrections to the tunnelling
rate by computing the bounce via the RG improved effective potential
in \eq{SMtreepotimproved} the result will look gauge dependent.  The
dependence on the gauge-fixing parameters in $\lambda_{\rm eff}$
(e.g.~in the Fermi gauge \cite{DiLuzio:2014bua}) is two-fold: it originates
from the fixed-order expression of the effective potential, and from
its running via the anomalous dimension of the field $h$.

From this point of view, gauge dependence (even if
numerically small) is a good thing, since is telling us that we are computing
something in the wrong way. 
As shown in Ref.~\cite{Isidori:2001bm}, 
the divergent corrections to the bounce action are formally gauge
independent, being directly related to the beta function of $\lambda$. 
A crucial role in order to achieve the cancellation of the gauge
dependent parts is played by the kinetic part of the effective action,
which is neglected altogether when dealing with the effective
potential only.

More generally, the gauge independence of the tunnelling
rate directly follows 
from the Nielsen identity \cite{Nielsen:1975fs,Fukuda:1975di}
\begin{equation}
\label{Nid}
\xi \frac{\partial S_{\rm eff}}{\partial \xi} 
= \int d^4 x \frac{\delta S_{\rm eff}}{\delta h(x)} K[h(x)], 
\end{equation}
where $\xi$ denotes the gauge-fixing parameter 
and $K$ is a functional of $h$ whose expression depends on the gauge fixing. 

The physical implication of \eq{Nid} is clear: the effective action is
gauge independent when evaluated on a configuration which extremizes
it. Hence, the bounce action is formally gauge independent as well.

The proof of the gauge independence of the tunnelling rate 
can be carried out in perturbation theory by means of a loop expansion of the
Nielsen identity, along the lines of Ref.~\cite{Metaxas:1995ab}. 
In practice, however, the cancellation of the gauge dependent parts 
in an explicit calculation might require some care. 
As recently observed in Refs.~\cite{Andreassen:2014eha,Andreassen:2014gha}, 
the usual loop expansion is not the consistent one for the SM, where 
$\lambda \sim \hbar$ (as in the original Coleman-Weinberg (CW) 
model \cite{Coleman:1973jx}) 
in order for the top-Yukawa corrections to destabilize
the tree-level electroweak vacuum. 
Consequently, such a modified loop expansion must be properly taken 
into account in order to observe the gauge independence of the SM tunnelling 
rate in perturbation theory. 
In particular, this entails the resummation of a particular class of daisy diagrams 
which (as observed in Ref.~\cite{Espinosa:2015qea}) is connected with the resummation of IR-divergent Goldstone loops \cite{Elias-Miro:2014pca,Martin:2014bca}.

Finally, we notice that for CW-like potentials, where the absolute minimum
is radiatively generated (as in the SM), the standard bounce formalism 
requires some modifications \cite{Weinberg:1992ds}. 
On the other hand, given the fact that for the measured values of the
SM parameters the lifetime of the electroweak vacuum turns out to be 
much larger than the age of the Universe, precision calculations of the SM 
tunnelling rate, although important, are not crucial at the moment.

\subsection{Tunnelling without barriers}
\label{tunnellingwb}

The approximation of taking $\lambda$ a negative constant might still
appear rather odd, since it corresponds to a tunnelling process from a
potential of the type $\lambda h^4$, with no barriers and a maximum in
$h = 0$.  It is well known, however, that the absence of a barrier in
the scalar potential is not necessarily a problem in field theory, due
to the presence of an extra barrier originated by the gradient
of the bounce~\cite{Lee:1985uv}. In this Section we want to explicitly
check this statement in the case of the SM.

The tunnelling process in field theory has to be understood as a
transition between two spatial field configurations at different
euclidean times $\tau_{i}$ and $\tau_{f}$
\beq
\lim_{\tau_i \rightarrow -\infty} h (\vec x, \tau_i) = v; \qquad
\lim_{\tau_f \rightarrow +\infty} h (\vec x, \tau_f) = v, 
\eeq
where $v$ denotes the false vacuum. The bounce action entering the
expression of the tunnelling probability in \eq{rate} can be recast in
a way that resembles the analogous one in quantum mechanics
\cite{Lee:1985uv}
\beq
\label{defSh}
S[h] = \int_{\tau_i}^{\tau_f} d\tau\, K(\tau) \sqrt{2 U[h]},   
\eeq
where the factor
\beq
\label{defKtau}
K(\tau) = \[ \int d^3 x \, 
\( \frac{\partial h}{\partial \tau} \)^2 \]^{\frac{1}{2}} 
\eeq
yields the correct normalization of the path length, and 
\beq
\label{defUh}
U[h] = \int d^3 x \, \[ \frac{1}{2} \( \vec\nabla h(\vec x,\tau) \)^2 
+ V (h(\vec x,\tau)) \]
\eeq
plays the role of the potential energy as in ordinary quantum mechanics.  

It is an instructive exercise to verify the existence of an actual
barrier in the case of the SM potential with $\lambda < 0$.  For
simplicity (and in order to proceed analytically) we take the mass
parameter $m=0$ in the scalar potential.
Starting from the $O(4)$-invariant bounce
solution \eq{SMh}, a straightforward calculation yields 
\begin{align}
&T[h] \equiv\int d^3 x \, \frac{1}{2} \( \vec\nabla h(\vec x,\tau) \)^2 
= \frac{2\pi^2}{\abs{\lambda}R}\(\frac{1}{1+\frac{\tau^2}{R^2}}\)^{\frac{3}{2}},
\\
&V[h] \equiv\int d^3 x \, V (h(\vec x,\tau))
 = - \frac{2\pi^2}{\abs{\lambda}R}
\(\frac{1}{1+\frac{\tau^2}{R^2}}\)^{\frac{5}{2}},
\end{align}
and
\beq
K(\tau) = \frac{2 \pi }{\sqrt{\abs{\lambda}R}} 
\frac{\frac{\tau}{R}}{\( 1 + \frac{\tau^2}{R^2} \)^{\frac{5}{4}}}.
\eeq
Following then the definitions in \eq{defUh} and \eq{defSh}, 
we finally get 
\beq
U[h] = \frac{2\pi^2}{\abs{\lambda}R} 
\(\frac{1}{1+\frac{\tau^2}{R^2}}\)^{\frac{3}{2}}
\[ 1 - \frac{1}{1+\frac{\tau^2}{R^2}} \],  
\eeq
and 
\beq
S[h] = 2 \int_0^\infty d\tau K(\tau) \sqrt{2 U[h]} 
= \frac{8 \pi^2}{3 \abs{\lambda}},   
\eeq
which reproduces the correct result for the SM bounce action. 

\begin{figure}[ht]
  \begin{center}
    \includegraphics[width=.45\textwidth]{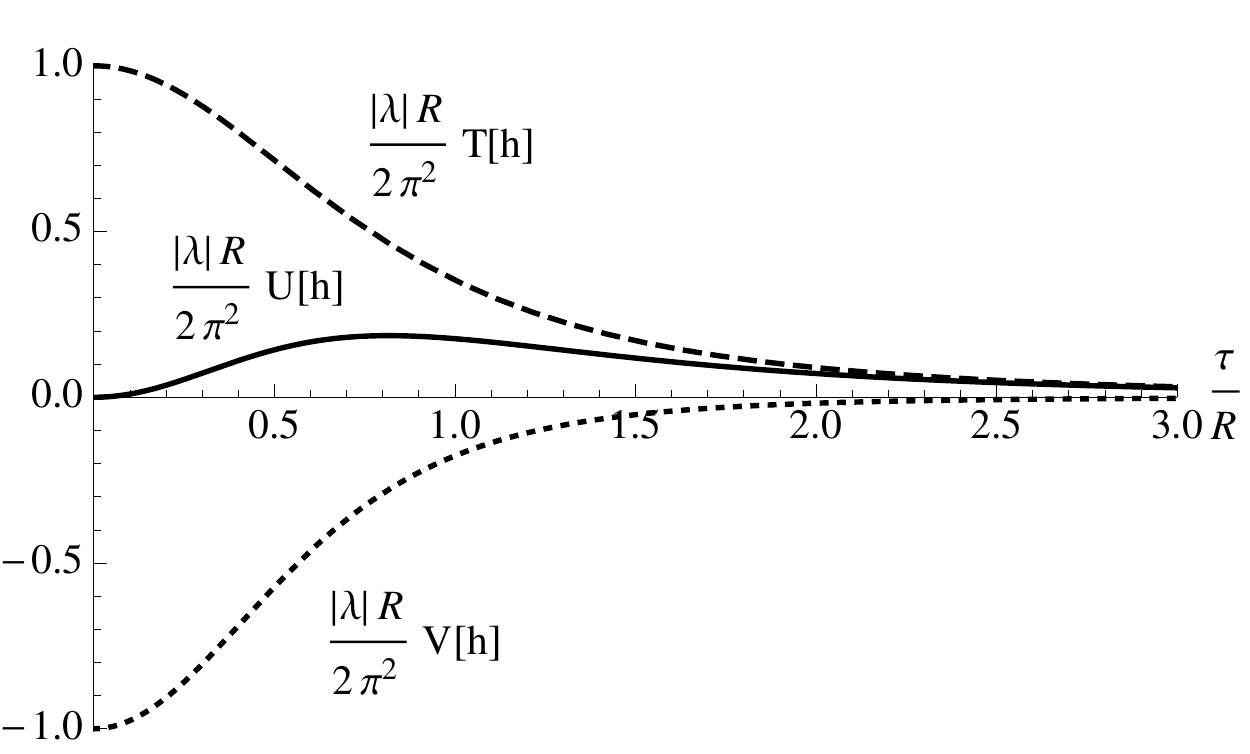}
  \end{center}
  \caption{\label{Barriers} Shapes of the gradient ($T[h]$) and
    potential ($V[h]$) barriers and their sum ($U[h]$) as a function
    of $\frac{\tau}{R}$.}
\end{figure}

The three quantities $T[h]$, $V[h]$ and $U[h]$ are plotted in 
\fig{Barriers}. 
$U[h]$ as a function of $\tau$ can be interpreted as the potential
energy along the path which minimizes the euclidean action. Thanks to
the positive gradient contribution $T[h]$ we see that there is a
barrier even for $\lambda$ constant and negative. Notice the correct
asymptotic behavior of $U[h]$, which tends to
zero as $\tau \rightarrow \infty$ 
since $h$ is approaching the false vacuum $v \rightarrow 0$. On
the other hand, $U[h] = 0$ for $\tau=0$ corresponds to complete barrier
penetration.  Since the point $\tau = 0$ coincides with $t =
0$ ($t = - i \tau$) 
the bounce solution can be analytically continued in
Minkowski space, so that the system evolves towards the true minimum
following the classical equation of motion.

From the above discussion one can also draw another important
conclusion: the largest energy scale relevant for barrier penetration 
is not that for which $V = 0$ (i.e.~the instability scale of
the SM effective potential), but rather the value of the bounce in its
center $h (\tau = 0)$ which corresponds to zero potential energy $U$.

\subsection{Violation of scale invariance by mass terms}
\label{mneq0}

In the previous sections (as in most of the existing literature on this
subject) the lifetime of the metastable vacuum of the SM was computed
neglecting the mass term of the Higgs boson, on the basis that the
electroweak scale, of order 10$^2$~GeV, is much smaller than $1/R$,
where $R$ is the typical size of the relevant bounce. We now wish to
discuss this approximation in some detail.

Let us consider the action
\beq
S[h]=\int d^4x\,\[\frac{1}{2}\partial_\mu h\partial_\mu h
+\frac{1}{2}m^2h^2+\frac{1}{4}\lambda h^4\],
\eeq
where $m^2>0$ and $\lambda<0$.
It was pointed out long ago~\cite{Affleck:1980mp}
that a bounce in this case does not exist.
The easiest way to see this is to perform a scale 
transformation, defined in Eq.~(\ref{scale}).
We get
\beq
S[h]\to S[h_a]=S[h]+\frac{m^2}{2a^2}\int d^4 x\,h^2(x),
\eeq
which cannot be stationary upon scale transformations unless
$h=0$:
\beq
\left.\frac{\partial S[h_a]}{\partial a}\right|_{a=1}=-m^2\int d^4 x\,h^2(x).
\label{nobounce}
\eeq
This phenomenon is well known in the context of studies of instanton gauge field
configurations.

Nevertheless, it is reasonable to think that, even in the presence of a mass
term, instanton configurations of the scalar fields should exist, provided
they are characterized by a length scale $R$ such that $m\ll 1/R$.
Roughly speaking, such a solution of the field equation is expected to
be a function of $r$, approximately constant for $r<R$ and approximately
zero for $r>R$. Furthermore, if no other mass scale is available,
the value of the bounce for $r<R$, $h(0)$, is proportional to
$1/R$ for dimensional reasons.
 Hence, the characteristic scale of $h$ may be identified
by the spacetime integral of a local operator, function of $h$.
For example,
\beq
\int d^4 x\,h^n(x)\sim h^n(0)\int_0^Rr^3 dr\sim R^{4-n}.
\eeq
Based on these intuitive considerations, it is suggested in 
Ref.~\cite{Affleck:1980mp} to perform a minimization of
the action functional in which the minimum configuration
is constrained to be characterized by a scale $R$ much smaller than 
$1/m$. The constraint is introduced by means of a suitable
Lagrange multiplier $\sigma$, i.e.~by adding to the action a term
\beq
S_c[h]=\sigma\[\int d^4x\,O(h)-cR^{4-n}\],
\label{constraint}
\eeq
where $O(h)$ is a local operator of mass dimension $n\neq 4$, for example
$O(h)=h^6$, and $c$ a constant. It is shown that with this modification an
instanton appears, called a {\it constrained instanton},
which differs from the bounce of the massless theory by powers
of $m^2 R^2$, times possibly logs of $mR$. Thus, the corrections to
the usually adopted approximation are indeed small, provided $mR\ll 1$.

The mechanism which restores the existence of a bounce in the massive
theory is illustrated in the Appendix of Ref.~\cite{Affleck:1980mp}
for the choice $O(h)=h^6$. The key point is
that the constraint Eq.~(\ref{constraint}), with $\sigma>0$, has the
effect of generating an absolute minimum (the true vacuum)
of the scalar potential,
which would be unbounded from below with $\sigma=0$ and 
$\lambda<0$.
Explicitly, the new scalar potential
\beq
V_c(h)=\frac{1}{2}m^2h^2+\frac{\lambda}{4}h^4 +\sigma h^6
\eeq
with $\lambda<0$, $m^2>0$ has a local minimum at $h=0$, with $V(0)=0$,
and an absolute minimum at $h\approx\sqrt{\frac{|\lambda|}{6\sigma}}$
(for $m^2\sigma\ll 1$). The presence of the constraining
term locally restores the scale invariance of the action:
\begin{align}
\left.\frac{\partial}{\partial a}\(S[h_a]+S_c[h_a]\)\right|_{a=1}
&=-m^2\int d^4 x\,h^2(x)+2\sigma\int d^4 x\,h^6(x)
\nonumber\\
&=-m^2\int d^4 x\,h^2(x)+\frac{2c\sigma}{R^2},
\end{align}
which is zero for
\beq
\sigma=\frac{m^2 R^2}{2c}\int d^4 x\,h^2(x)\sim (mR)^2 R^2\ll R^2.
\eeq

This issue is not directly relevant in the SM; indeed,
as we have seen in Sect.~\ref{violscalradiative}, 
scaling violation induced by radiative
corrections have the effect of selecting a bounce of size
$R_{\rm SM}\sim 10^{-17}$~GeV$^{-1}$. Explicitly, with
\beq
V(h)=\frac{1}{4}\lambda(h)h^4
\eeq
we have
\beq
\left.\frac{\partial}{\partial a}S[h_a]\right|_{a=1}
=\frac{1}{4}\int d^4 x\,\beta_\lambda(h)h^4(x),
\eeq
and $\beta_\lambda(\mu)$ is zero around $\mu=1/R_{\rm SM}\sim
10^{17}$~GeV.  As a consequence, scale invariance is locally restored,
and a bounce of size $R\sim R_{\rm SM}$ is found. The effects of
quadratic and cubic terms in the potential are suppressed by powers of
$vR_{\rm SM}\sim 10^{-15}$ and can be safely neglected. Nevertheless,
the simple example discussed above shows that the explicit violation
of scale invariance (for example, by the introduction of
non-renormalizable terms in the scalar potential) may lead to
complications in the calculation of the vacuum tunnelling rate.

\section{Violation of scale invariance by new interactions}
\label{sec:nonren}

We now turn to a discussion of the effects of SM extensions 
characterized by the appearance of an explicit 
new energy scale in the effective potential. 
Following Refs.~\cite{Branchina:2013jra,Branchina:2014usa,Branchina:2014rva},
we describe the effect of generic NP occurring at high-energy scales
by introducing two non-renormalizable terms in the scalar potential, 
\beq
V(h)=\frac{1}{4}\lambda h^4+\frac{\lambda_6}{6M^2}h^6
+\frac{\lambda_8}{8M^4}h^8.
\label{newpot0}
\eeq
Assuming the new couplings $(\lambda_{6,8}$) to be of order one, the
extra terms affect the potential only for field values of order $M$.
The scalar potential in Eq.~(\ref{newpot0}) should be understood just
as a toy model with no claim of being realistic. It is however
sufficient in order to highlight some basic features of the tunnelling
rate in the presence of new interactions.  A more realistic expression
of the modified potential in presence of NP is obtained by considering
a full tower of higher-dimensional operators, namely
\beq
V(h)=\frac{1}{2}\lambda_2 M^2 h^2+\frac{1}{4}\lambda h^4+\sum_{n=3}^\infty
\frac{\lambda_{2n}}{2nM^{2n-4}}h^{2n},
\label{newpot}
\eeq
where we have also inserted a mass term due to the presence of a physical 
threshold at the scale $M$,\footnote{We thank Ulrich Ellwanger for pointing out this effect to us.} assuming that, by a suitable fine-tuning,
the electroweak scale is kept at the correct phenomenological 
value.

Most of the following discussion and results apply to the general
potential in Eq.~(\ref{newpot}), the key point being the explicit
breaking of scale invariance characterized by the scale
$M$.\footnote{The whole discussion applies as well to the case where
  such effective interactions arise within an explicit renormalizable
  model with new degrees of freedom, such as the one proposed in
  Ref.~\cite{Branchina:2015nda}.}  
In principle, effective operators
involving derivatives of the Higgs field should also be included,
since they influence the determination of the bounce as well.  By
employing a suitable field redefinition, it can be
shown~\cite{Isidori:2007vm} that they can affect the bounce only when
their dimension $d$ exceeds $d=6$.

In Refs.~\cite{Branchina:2013jra,Branchina:2014rva,Branchina:2014usa}
the mass scale $M$, required by the presence of non-renormalizable terms,
is taken to be of the order of the Planck mass. 
Here we postpone for a while the choice of a definite value for $M$, 
keeping the discussion on general terms. 
We will however be especially interested in the case $M \gg 1/R_{\rm SM}$, 
since for $M \lesssim 1/R_{\rm SM}$ the SM bounce gets trivially modified and 
the lifetime of the electroweak vacuum can be any with respect to the SM case.

Because of the dependence on $M$ of the effective potential,
the action is no longer scale invariant. Hence, 
the existence of a bounce for generic choices of the new couplings
$\lambda_{2n}$ is no longer guaranteed. 
For example, with all $\lambda_{2n}=0$ except $\lambda_6$ the action
can only be stationary upon scale transformations if $h=0$. We have indeed
\beq
S[h]\to S[h_a]=S[h]+\frac{\lambda_6a^2}{6M^2}\int d^4 x\,h^6(x),
\eeq
which implies
\beq
\left.\frac{\partial S[h_a]}{\partial a}\right|_{a=1}
=\frac{\lambda_6}{3M^2}\int d^4 x\,h^6(x).
\label{nobounce6}
\eeq 
However, in analogy with the case of a mass term discussed in
Section~\ref{mneq0}, one may argue that a constrained instanton with
$R \gg 1/M$ may still exist.  Furthermore, if more non-renormalizable
couplings are different from zero (e.g.~both $\lambda_6\not=0$ and
$\lambda_8\not=0$), the r.h.s.~of \eq{nobounce6} contains more terms.
In such case one can conceive the possibility that these terms
compensate each other (assuming some of the couplings have opposite
signs) for instanton solutions characterized by $R \sim 1/M$.  On
general grounds, we thus expect two classes of instanton solutions
(when they exist): those characterized by $R \gg 1/M$ and those
characterized by $R \sim 1/M$.

With a generic potential of the type of \eq{newpot}, 
the field equation cannot be
integrated analytically, and one has to adopt numerical methods (which
we are going to do below). However, a somewhat deeper understanding of
the tunnelling process in the presence of NP can be achieved by making
contact with the case of the pure SM. To this purpose, it is useful to
rewrite the modified potential \eq{newpot} in terms of an effective
quartic coupling:\footnote{Note that this parametrization 
is not directly related to the effective potential.}
\begin{equation}
\label{scalpotNPrewr}
V(h) = \frac{1}{4} \lambda_{\rm{eff}}(h) h^4, 
\end{equation}
with 
\begin{equation}
\label{deflameffB}
\lambda_{\rm{eff}} (h) = \frac{2\lambda_2 M^2}{h^2}+\lambda + 4\sum_{n=3}^{\infty} 
\frac{\lambda_{2n}}{2nM^{2n-4}}h^{2n-4}, 
\end{equation}
where, unless otherwise specified, we take the quartic coupling $\lambda$ to be 
at its minimal negative value (as in the SM case) and we neglect the 
logarithmic running for the NP couplings, since 
their are anyway unknown. Note that the coupling $\lambda_2$ becomes active only above threshold $h \gtrsim M$. 

If we were able to argue that the dominant contribution to the bounce action
is provided by taking the argument of $\lambda_{\rm eff}$ at some
fixed value, then the action would be approximately given by the SM expression
\eq{BactionSM}, with $\lambda$ replaced by the appropriate value of
$\lambda_{\rm eff}$.

It can be seen that this is indeed the case by adopting the
approximation mentioned in Section~\ref{mneq0}, namely, to consider
the bounce as a finite-action, $O(4)$-invariant solution of the
euclidean field equation $h(r)$, approximately constant for $r<R$ and
approximately zero for $r>R$.  Let us first test this approximation in
the case of the SM with no mass term, where the exact
result is known. In this case, the constant value $h(0)$ of the bounce
for $0<r<R$ must be proportional to $1/(R\sqrt{|\lambda|})$.  The
factor $1/R$ arises for dimensional reasons.  The factor
$1/\sqrt{|\lambda|}$ can be understood by writing the action as
\begin{align}
S_{\rm SM}[h]=&\int d^4x\,\[\frac{1}{2}\partial_\mu h\partial_\mu h
+\frac{1}{4}\lambda h^4\]
\nonumber\\
=&\frac{1}{\lambda}\int d^4x\,\[\frac{1}{2}\partial_\mu H\partial_\mu H
+\frac{1}{4}H^4\]
\end{align}
where the bounce
\beq
H(x)=\sqrt{\lambda} h(x)
\eeq
does not depend on $\lambda$. 

Within this approximation we get
\begin{align}
S_{\rm SM}[h]=&2\pi^2\int_0^{\infty}dr\,r^3\[V(h)-\frac{1}{2}h(r)V'(h)\] 
\nonumber\\
\approx&\frac{\pi^2R^4}{2}\[V\(h(0)\)-\frac{1}{2}h(0)V'\(h(0)\)\] 
\nonumber\\
=&\frac{\pi^2R^4}{8}|\lambda|h^4(0)
\nonumber\\
=&\frac{a^4\pi^2}{8|\lambda|},
\label{approxactionSM}
\end{align}
where $a$ is a constant, and
we have used the field equation in the first equality.
The correct value $S_{\rm SM}[h]=8\pi^2/(3|\lambda|)$ is reproduced
for $a^2=8/\sqrt{3}\approx (2.15)^2$.

We now turn to the case of the modified potential
\eq{newpot}. We observe that, within the same
approximation that led us to \eq{approxactionSM}, 
the effect of a scale transformation \eq{scale} on the
modified action is
\beq
\left.\frac{\partial S[h_a]}{\partial a}\right|_{a=1}
=\int d^4 x\,\frac{1}{4}\frac{\partial\lambda_{\rm{eff}}(h)}{\partial h} h^5(x)
\sim h(0)\lambda'_{\rm{eff}}\(h(0)\).
\label{scalvarlameff}
\eeq
Hence, in order for the action to be stationary under scale transformations, 
$h(0)$ should be chosen so that
\beq
\lambda'_{\rm{eff}}\(h(0)\) = 0
\eeq 
locally restores scale invariance in a neighbourhood of $h(0)$
(the case $h(0)=0$ is obviously uninteresting).
If such a value of $h(0)$ exists, then
\begin{align}
\label{approxaction}
S[h]=&2\pi^2\int_0^{\infty}dr\,r^3\[V(h)-\frac{1}{2}h(r)V'(h)\] 
\nonumber\\
\approx&\frac{\pi^2R^4}{2}\[V\(h(0)\)-\frac{1}{2}h(0)V'\(h(0)\)\] 
\nonumber\\
=&\frac{\pi^2R^4}{2}\[-\frac{1}{4}\lambda_{\rm eff}\(h(0)\)h^4(0)
-\frac{1}{8}h^5(0)\lambda'_{\rm eff}\(h(0)\)\]
\nonumber\\
=&-\frac{\pi^2R^4}{8}\lambda_{\rm eff}\(h(0)\)h^4(0),
\end{align}
which is precisely what we would get in the SM with no
mass term, and $\lambda$ replaced by $\lambda_{\rm eff}\(h(0)\)$,
see the third line in \eq{approxactionSM}.
We conclude that the leading contribution to the bounce action is given by
\beq
\label{BactionSMeff}
S[h] \approx \frac{8\pi^2}{3|\lambda_{\rm{eff}}\(h(0)\)|}, 
\eeq
provided
\beq
h(0)=\frac{a}{R\sqrt{|\lambda_{\rm eff}\(h(0)\)|}};\qquad
a=\sqrt{\frac{8}{\sqrt{3}}}.
\label{h0}
\eeq
As an example, let us consider the potential in \eq{newpot0}. We have 
in this case
\begin{align}
\lambda_{\rm{eff}}(h) &=\lambda+\frac{2}{3}\lambda_6\frac{h^2}{M^2}
+\frac{1}{2}\lambda_8 \frac{h^4}{M^4},
\label{lameffzero}\\
\lambda'_{\rm{eff}} (h) &=
\frac{2}{3}\frac{h}{M^2}\left(2\lambda_6 + 3\lambda_8 \frac{h^2}{M^2} \right).
\label{lampeffzero}
\end{align}
We can distinguish three basic  scenarios.
\begin{itemize}
\item[{\bf I.}] 
One possibility is that both $\lambda_6$ and 
$\lambda_8$ are non-negative and at least one of them 
is different from zero. In such case the potential is bounded
from below and, for any 
value of $h$, $\lambda_{\rm eff} (h) >  \lambda (h)$.

In the limit where we assume $\lambda=-|\lambda|$ constant, 
$\lambda'_{\rm{eff}}(h)$ is always different from zero: 
local scale invariance is hopelessly lost, and a bounce cannot be found.
These expectations are confirmed by our numerical analysis, described
in~\ref{numerical}. Actually one finds that a bounce with a SM-like action 
exists for finite values of $x_{\rm max}$, but its
size $R$ grows to infinity as $x_{\rm max}$ is sent to infinity.
Since $h(0)$ is inversely proportional to $R$, it follows 
that the bounce is in fact zero everywhere.

However, the absence of a bounce turns out to be an artefact 
of choosing $\lambda$ constant in \eq{deflameffB}. 
With $\lambda$ running there is an extra compensating 
contribution in \eq{lampeffzero}, and indeed a bounce of finite size
with a SM-like bounce action can be found numerically, 
provided $M\gg 1/{R_{\rm SM}} \approx 10^{17}$ GeV, the scale at which
$\lambda$ has a minimum.
Hence we conclude that in such a case the decay probability of the false vacuum is not modified with respect to the SM value.

\item[{\bf II.}]
A second possibility is that $\lambda_6$ and $\lambda_8$ have opposite signs,
but the potential remains bounded from below, 
hence  $\lambda_6 < 0$ and $\lambda_8 > 0$.
In this case, the two non-renormalizable terms compensate each other
and restore scale invariance locally (in the neighborhood of 
field configurations characterized the scale $1/M$).
Indeed, $\lambda'_{\rm{eff}}(h)$ is zero for
\beq 
h=h(0)=M\sqrt{\frac{2|\lambda_6|}{3\lambda_8}},
\eeq
and
\beq
\lambda_{\rm{eff}} \(h(0)\)
=-|\lambda|-\frac{2|\lambda_6|^2}{9\lambda_8}.
\eeq
The instanton size is now immediately read off \eq{h0}:
\beq
R=\frac{a}{h(0)\sqrt{|\lambda_{\rm eff}\(h(0)\)|}}
=\frac{a}{M}\frac{1}
{\sqrt{\frac{2|\lambda_6|}{3\lambda_8} 
\(|\lambda|+\frac{2|\lambda_6|^2}{9\lambda_8}\)}}\sim\frac{1}{M}.
\eeq
Two important points are to be noted. First, we have in this case
\beq
|\lambda_{\rm{eff}}\(h(0)\)|
=|\lambda|+\frac{2|\lambda_6|^2}{9\lambda_8}>|\lambda|,
\eeq
and therefore
\beq
S[h]<S_{\rm SM}[h].
\eeq
As a consequence, the decay probability
of the false vacuum can only be increased with respect to the pure SM.
Second, the dominant contribution to the tunnelling rate is independent
of the scale $M$ of NP, because the action is. This conclusion 
has been reached in the context of an approximate calculation, but it is
easy to see that it holds true in general. Indeed,
one may define dimensionless coordinates and field by
\beq
x=Mr;\qquad \tilde h(x)=\frac{h(r)}{M},
\eeq
so that
\beq
V(h)=
M^4\[\frac{1}{4}\lambda \tilde h^4+\sum_{n=3}^\infty
\frac{\lambda_{2n}}{2n}\tilde h^{2n}\]\equiv M^4\tilde V(\tilde h).
\eeq
Then
\begin{multline}
\label{BactionM}
S[h]=2\pi^2\int_0^{\infty}dr\,r^3\[V(h)-\frac{1}{2}h(r)V'(h)\right] \\
=2\pi^2\int_0^{\infty}dx\,x^3\[\tilde V(\tilde h)
-\frac{1}{2}\tilde h(x)\tilde V'(\tilde h)\right]
\end{multline}
which is independent of $M$.
This is an important point: whatever modification may the potential in
Eq.~(\ref{newpot}) induce on the tunnelling rate, this is independent
of the scale $M$ up to subleading pre-exponential corrections. 

\item[{\bf III.}]  The last case to be considered is the one where
  the potential is unbounded from below at large field values.  When both
  $\lambda_6$ and $\lambda_8$ are negative (or at least one is
  negative and the other is zero), it is clear that the bounce does
  not exist because of the lost of scale invariance.

The case $\lambda_6 > 0 $ and $\lambda_8 <0 $ is more subtle: here a
non-zero solution of $\lambda'_{\rm eff}(h)=0$ does exist, but it
corresponds to a maximum of $\lambda_{\rm eff}$ and not to a bounce
configuration. Actually the fast drop of the potential at large field
values makes the whole problem of finding a bounce solution
ill-defined in this case: there exist ``rolling solutions" that destabilize 
the electroweak vacuum characterized by a much 
shorter time scale~\cite{Lee:1985uv}.
\end{itemize}

\begin{table}[t]
  \centering
  \begin{tabular}{@{} |c|c|c|c|c|c| @{}}
  \hline
     $\lambda$ & $\lambda_6$ & $\lambda_8$ & $h$ & $R$ & $S[h]$ \\ 
 \hline
    const & $<0$ & $<0$ & 0 & $\infty$ & 0 \\
    const & $<0$ & $>0$ & $\neq 0$ & $\sim 1/M$ & $< S_{\rm SM}[h]$ \\
    const & $>0$ & $<0$ & 0 & $\infty$ & 0 \\
    const & $>0$ & $>0$ & 0 & $\infty$ & 0 \\
    \hline
    run & $<0$ & $<0$ & 0 & $\infty$ & 0 \\
    run & $<0$ & $>0$ & $\neq 0$ & $\sim 1/M$ & $< S_{\rm SM}[h]$ \\
    run & $>0$ & $<0$ & 0 &  $\infty$ & 0 \\
    run & $>0$ & $>0$ & $\neq 0$ & $R_{\rm SM}$ & $S_{\rm SM}[h]$ \\
    \hline
  \end{tabular}
  \caption{\label{summarybounce} Summary of the bounce solutions 
  associated with the potential in \eq{newpot0}. Only the case $M \gg 1/R_{\rm SM}$ is considered, 
  while for $M \lesssim 1/R_{\rm SM}$ the bounce action can be any with respect to the SM one.}
\end{table}

\begin{figure}[t]
  \begin{center}
    \includegraphics[width=.45\textwidth]{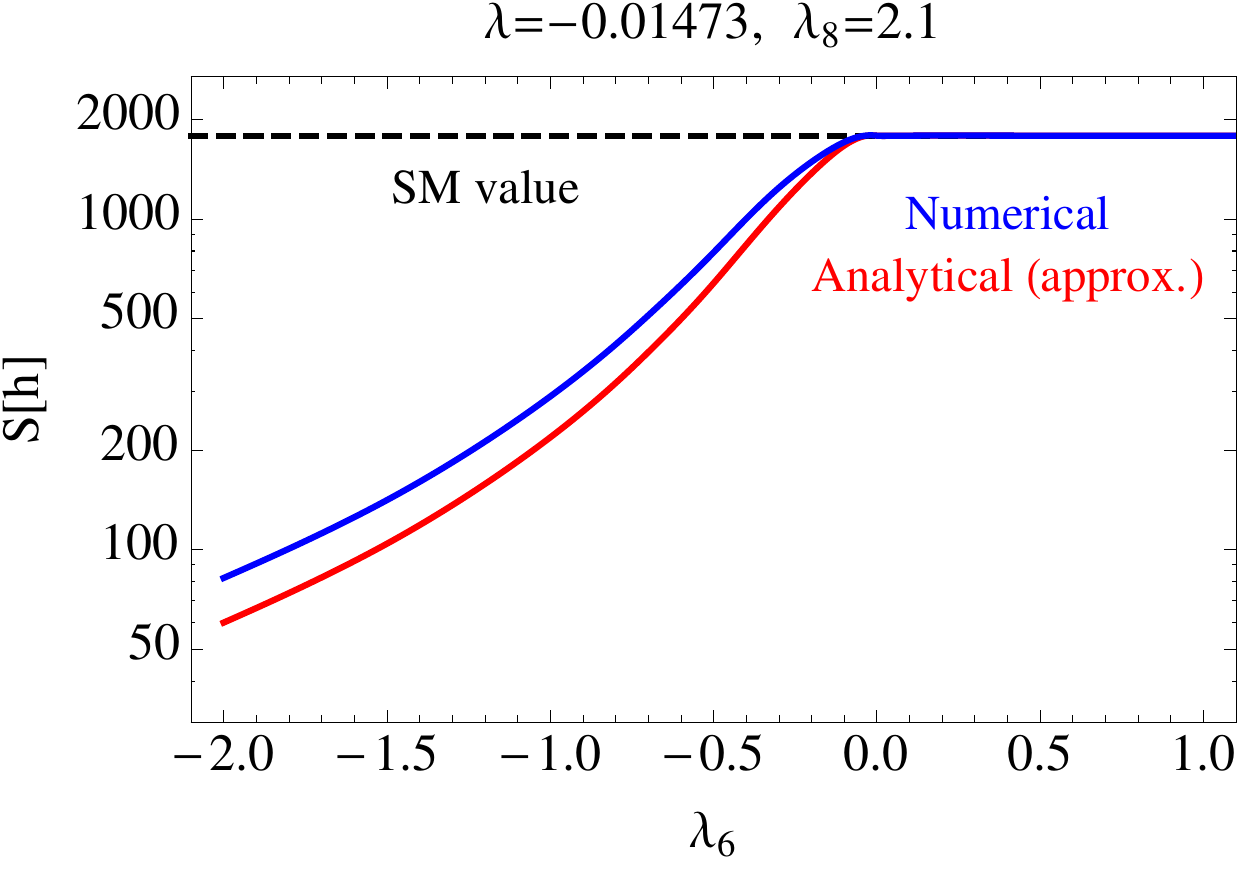}
  \end{center}
\caption{\label{Bnumvsan} Numerical vs.~analytical (approximated)
  determination of the bounce action as a function of $\lambda_6$. The
  relative difference is always within $30 \%$.  Formally, for
  $\lambda$ constant and $\lambda_6 > 0$ a bounce does not exist;
  the result in the plot for $\lambda_6 > 0$ has to be understood for
  $\lambda$ running.}
\end{figure}

\begin{figure}[t]
  \begin{center}
    \includegraphics[width=.45\textwidth]{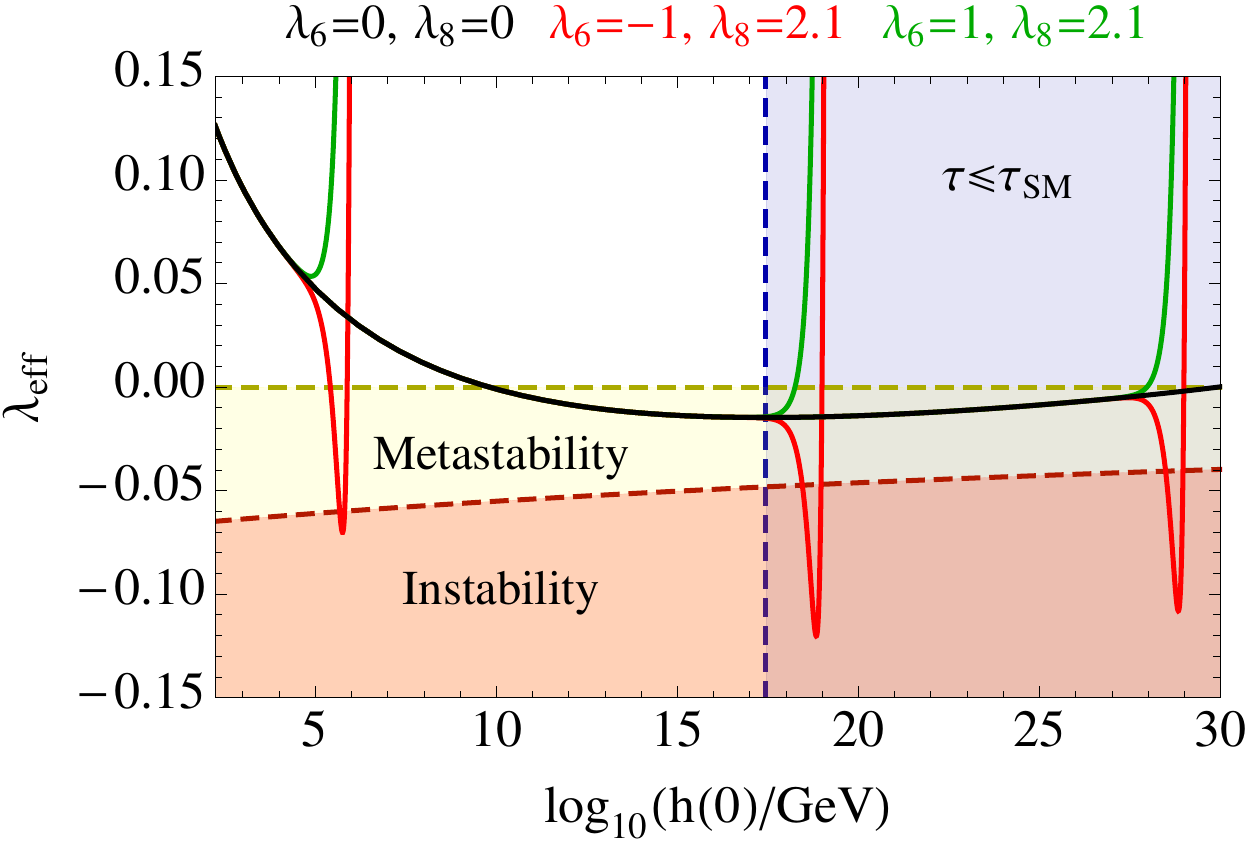}~~ \\
  \end{center}
  \caption{\label{lameffNP} 
   Evolution of $\lambda_{\rm eff} (h(0))$ (including the running of $\lambda$) for three
    different values of the NP scale, 
    $M=10^6~\text{GeV}, M_P$ and $10^{10} M_P$,
    and for three different choices of the NP parameters $\lambda_{6,8}$. 
    The instability bound is defined by the 
    inequality $\abs{\lambda_{\rm eff}(h(0))}>\frac{2\pi^2}{3}\frac{1}{\log{(\tau_U h(0))}}$, 
    where $\tau_U = 4.35 \times 10^{17}$ sec is the age of the Universe.}
\end{figure}

These results are summarized in \Table{summarybounce}, with either
constant or running $\lambda$.\footnote{A similar study of the
tunnelling rate as a function of $\lambda_6$ and $\lambda_8$ was
performed in Ref.~\cite{Lalak:2014qua}.} To complete this study,
we show in \fig{Bnumvsan} 
the comparison between the numerical determination of the bounce
action and the analytical approximation in \eq{BactionSMeff}. 
The agreement is generally quite good. 
Hence, one can take advantage of the $\lambda_{\rm eff}$ language 
in order to describe in an essential way the tunnelling process, as exemplified in \fig{lameffNP}, 
where some potentials corresponding to cases I and II are considered. 
From that it is clear that modifications of the SM potential above $1/R_{\rm SM} \approx 10^{17}$ GeV 
may only shorten the lifetime of the electroweak vacuum.

The above discussion can be generalized to the case of more
non-renormalizable terms, including derivative operators, the three
main categories being defined by the stability (I and II) or
instability (III) of the potential at large field values, and by the
presence of a new minimum around the scale $M$ (case II).

Summarizing, the explicit breaking of scale invariance at energy scales 
above $1/R_{\rm SM} \approx 10^{17}$ GeV may lead to a new 
bounce, whose characteristic scale is necessarily related to 
that of NP (case II). However, this happens only if the NP 
modifies the ground state of the theory, generating a new deep minimum
around the new scale $M$. In such case, 
the effect of NP is that of opening a new decay channel for the 
electroweak vacuum and the tunnelling rate can 
only increase with respect to the SM calculation. It is then obvious that in
this case one is not analyzing anymore the stability of the SM, but  that of a
different theory, with a completely different ground state.

The above argument also explains why there is an apparent violation of  
the decoupling theorem, according to which one  expects
new-degrees of freedom characterized by a scale $M \gg 1/R_{\rm SM}$ 
to be irrelevant for the analysis of the vacuum stability. 
This apparent paradox is due to the fact that the modifications of the 
effective potential introduced by means of Eq.~(\ref{newpot0})
are not only related to the appearance of high-frequency 
dynamical modes (for which the decoupling still applies), 
but they also imply  a drastic modification of the ground state 
of the theory (invalidating the decoupling argument).

We finally note that the modified effective potential of the type in Eq.~(\ref{newpot0}),
leading to a fast tunnelling rate (case II), 
is not a well motivated UV completion of the SM potential close to the Planck scale. 
On the one hand, except for fine-tuned values of  $\lambda_6$ and $\lambda_8$,
the lifetime  of the electroweak vacuum turns out to be extremely fast,
in sharp contradiction with the existence of the present Universe. This implies that 
such modified potential cannot be considered as a phenomenologically 
viable UV completion of the SM potential.  On the other hand, 
if $M$ is close to the Planck mass,   
truly gravitational effects~\cite{Coleman:1980aw} cannot be ignored.
As recently pointed out in Ref.~\cite{Espinosa:2015zoa}, the latter 
suppress the tunnelling rate by many orders of 
magnitude.\footnote{Gravitational effects in minimal Einstein gravity tend to slow down the tunnelling rate \cite{Coleman:1980aw}. 
However, as long as $1/R_{\rm SM} \ll M_P$, these corrections are small 
and hence the presence of gravity does not inficiate the results obtained within the SM in isolation \cite{Isidori:2007vm}.}

\section{Conclusions}
\label{concl}

The decay probability of the electroweak vacuum under quantum
tunnelling is intimately related to the breaking of scale invariance
in the Higgs effective potential. The latter can be characterized by
an effective $h^4$ interaction, both within and beyond the SM, which is
not scale invariant beyond the tree level.  To a good
approximation, the size of the bounce and the tunnelling probability
are determined by the energy scale at which this effective coupling,
$\lambda_{\rm eff}(h)$, reaches its minimum value and by its value at
the minimum, respectively.  In the absence of NP, the breaking of
scale invariance occurs dominantly via SM radiative corrections. The
latter selects a leading bounce characterized by the scale $1/R_{\rm
  SM} \approx 10^{17}$ GeV, that implies a sufficiently long lifetime
of the electroweak vacuum compared to the age of the Universe.

New degrees of freedom at high energies can, in principle, 
introduce a new explicit breaking of scale invariance in the 
effective potential. This, in turn, can shift the energy scale 
where $\lambda_{\rm{eff}} (h)$ reaches its minimum value and, 
most importantly, can modify such minimum value. 
If the new degrees of freedom appear above the scale $1/R_{\rm SM}$, 
we can distinguish two cases: those where NP simply stabilizes the SM 
potential, and those where NP introduces new decay channels for the 
electroweak vacuum. In the first case, characterized by the absence of
new minima for $\lambda_{\rm{eff}} (h)$, the lifetime of the electroweak vacuum 
remains unchanged. In the latter case it  can only decrease
with respect to the pure SM. In particular, a shorter lifetime occurs only 
if the new degrees of freedom drastically change the ground-state of the theory,
introducing a new minimum for  $\lambda_{\rm{eff}} (h)$, 
such that it does not make sense anymore to speak about the stability problem 
of the SM potential. 
 
Given these arguments, we can conclude that the sensitivity
of the tunnelling rate to the possible UV completions of the model 
does not invalidate the vacuum stability
analyses performed using the pure SM potential. The scope of the latter is answering  the following 
well-defined physical question: {\em Does the extrapolation of the SM up to the Planck scale 
necessarily implies  the existence of NP below such scale?}  According to the present experimental values of $m_h$ and $m_t$,
we can state that the answer to this question is {\em no}~\cite{Degrassi:2012ry,Buttazzo:2013uya}.
Were the top mass e.g.~$180$ GeV, the answer would have been different since 
NP would have been necessarily implied \emph{well below} the scale where $\lambda$ reaches its minimum SM value, 
regardless of the physics in the deep UV which \emph{cannot} improve on stability.  
We finally stress that the negative answer to the above question does not necessarily imply the 
absence of NP up to the Planck scale: it only says that the model is compatible with the absence of 
NP up to the Planck scale. Similarly, from this analysis we cannot infer that any UV completion 
of the theory at the Planck scale is compatible with the stability of the electroweak vacuum, 
but only that it is possible to build UV completions that do not contain new degrees of
freedom below the Planck scale.

\section*{Acknowledgments}

We thank Jos\'e R.~Espinosa and Alessandro Strumia for reading the
manuscript and for useful comments.
G.R. thanks Camillo Imbimbo for useful discussions. The
work of L.D.L.~is supported by the Marie Curie CIG program, project
number PCIG13-GA-2013-618439. The work of G.I. is supported in part by the Swiss National Science Foundation 
(SNF) under contract 200021-159720. The work of G.R. is supported in part by
an Italian PRIN2010 grant.

G.R. wishes to dedicate this work to Maurizio Lo Vetere.
\appendix
\section{Numerical determination of the bounce}
\label{numerical}

For the numerical analysis it turns out to be convenient 
to work with adimensional variables 
\beq
\label{cofv}
x=Mr;\qquad \tilde h(x)=\frac{h(r)}{M};\qquad
\tilde V(\tilde h)=\frac{V(h)}{M^4},
\eeq
where $M$ is an arbitrary mass scale. The bounce equation reads
\beq
\tilde h''(x)+\frac{3}{x}\tilde h'(x)=\tilde V'(\tilde h),
\label{bounceadim}
\eeq
with boundary conditions 
\begin{align}
\label{BC1}
&\lim_{x \rightarrow \infty} \tilde h(x) = v / M, \\ 
\label{BC2}
&\left. \frac{d \tilde h(x)}{d x} \right|_{x = 0} = 0.
\end{align}
A numerical solution can be found by means of the shooting method 
(see also \cite{Dunne:2005rt,Lalak:2014qua,Branchina:2014rva}), 
which consists in tuning the initial condition $a_0$ 
in such a way that \eq{BC1} is satisfied at the boundary.  
In practice, since we cannot start at the singular point $x=0$, 
we Taylor-expand the solution around the origin and define the Cauchy problem 
\begin{align}
\tilde h (x_{\rm min}) &= a_0 + \frac{1}{8} \tilde V'\(h(0)\) x_{\rm min}^2, \\
\tilde h'(x_{\rm min}) &= \frac{1}{4} \tilde V'\(h(0)\) x_{\rm min}, 
\end{align}
where we kept up to quadratic terms in $x_{\rm min}$.  
By evolving the solution $\tilde h (x; a_0)$ up to $x = x_{\rm max}$, 
we require
\begin{equation}
\tilde h (x_{\rm max}; \bar a_0) = 0,  
\end{equation}
which determines the value of $\bar a_0$ which satisfies the boundary 
condition at $x = x_{\rm max}$. 

Analogously, by Taylor-expanding the solution at the first non-trivial order 
in $1/x_{\rm max}$ for large $x_{\rm max}$, 
we get ($m = 0$ case)
\begin{equation}
\tilde h (x_{\rm max}) = \frac{a_{\infty}}{x^2_{\rm max}}, 
\end{equation}
which is useful in order to have an analytical control over the asymptotic
solution. 
On the other hand, if the mass term is kept in the potential ($m \neq 0$ case) 
the asymptotic solution is (see e.g.~Appendix of \cite{Nielsen:1999vq})
\begin{equation}
\tilde h (x_{\rm max}) = a_{\infty} \frac{\epsilon}{x_{\rm max}} 
K_1(\epsilon x_{\rm max}) 
\approx a_{\infty} \frac{\epsilon}{x_{\rm max}} 
\sqrt{\frac{\pi}{\epsilon x_{\rm max}}} e^{-\epsilon x_{\rm max}}, 
\end{equation}
where $\epsilon = m/M$
and in the last step we used the asymptotic expression of
the modified Bessel function $K_1(x)$ for large $x$. 

The bounce action is determined by 
integrating the profile of the bounce $h (x; \bar a_0)$ 
between $x_{\rm min}$ and $x_{\rm max}$
\beq
\label{BAadim}
S[h] = \frac{2 \pi^2}{M^4} \int_{x_{\rm min}}^{x_{\rm max}} 
x^3 dx \, \left[ V (\tilde h M) - \frac{1}{2} \tilde h M V' (\tilde h M) \right], 
\eeq
where we applied the equation of motion 
and performed the change of variables in \eq{cofv}.
Finally, the algorithm is iterated by choosing 
increasingly smaller (larger) values of $x_{\rm min}$ ($x_{\rm max}$).

\bibliographystyle{elsarticle-num} 
\bibliography{bibliography}

\end{document}